\documentstyle[amssymb,12pt]{article}
\begin{document}
\author{Andrea Gamba\footnote{Dipartimento di Matematica, Politecnico di Torino,
10129 Torino, Italy, and INFN, Sezione di Milano, 20133 Milan, Italy;
e-mail: gamba@polito.it.}\ \
and Igor V. Kolokolov\footnote{Budker Institute of Nuclear Physics,
Novosibirsk 630090, Russia;
e-mail: kolokolov@inp.nsk.su.}
}

\title{Dissipation statistics of a passive scalar in a multidimensional
smooth flow
\footnote{Submitted to J. Stat. Phys.}
}

\date{\ }

\maketitle

\begin{abstract}

We compute analytically the probability distribution function
${\cal P}(\varepsilon)$ of the dissipation field \( \varepsilon =(\nabla \theta )^{2} \)
of a passive scalar \( \theta  \) advected
by a \( d \)-dimensional random flow, in the limit of large Peclet and Prandtl
numbers (Batchelor-Kraichnan regime). The tail of the distribution
is a stretched exponential: for \( \varepsilon \rightarrow \infty  \), \( \ln {\cal P}(\varepsilon )\sim -(d^2\varepsilon )^{1/3} \).

\end{abstract}

\begin{center}
{\ 
\footnotesize turbulence; intermittency;
functional integral}
\end{center}

\section{Introduction}
\label{introduction}
Intermittency and strong non-gaussianity of developed turbulence are
most clearly reflected in the peculiar structure of the observed probability
distribution functions (p.d.f.) of the gradients of the turbulent field.
A typical logarithmic plot of the gradients p.d.f. is concave rather than
convex, showing a strong central peak and slowly decaying tails.
Rare strong fluctuations are responsible for the tails, while large
quiet regions are related to the central peak.
In particular, such p.d.f. were observed for the square gradients
\( \varepsilon=(\nabla\theta)^2 \) (dissipation field) of a
scalar field \( \theta \) passively advected by an incompressible
turbulent flow.

From the theoretical standpoint, most rigorous results in the theory
of turbulent mixing have been obtained so far within the framework
of the Kraichnan model \cite{batchelor59,kraichnan74},
describing the advection of
a passive scalar by a random velocity field, delta-correlated in time.
The exact solvability of the Kraichnan model suggests that it can play
in turbulence a role similar to that played by the Ising model
in the theory of critical phenomena~\footnote{The similarity should
be understood in a loose sense. The Kraichnan model can be viewed in
itself as a sort of mean-field approximation to a full theory of
developed turbulence \cite{fl94}.}.

In particular, the p.d.f. of the dissipation field \(\varepsilon\)
was recently calculated analytically in the one-dimensional \cite{ckv97P}
and two-dimensional
\cite{cfk98} cases.
In this paper we extend the result of Ref. \cite{cfk98} to
arbitrary space dimensions \(d\).
Our techniques are based on a combination of functional integration
and group-theoretical methods, first introduced by one of us
\cite{kolokolov86}
in the context of quantum magnetism.
These techniques allowed first to compute exactly the p.d.f.
of the passive scalar in the two-dimensional, non-dissipative
case \cite{cfkl95}.
The multidimensional case was considered in
\cite{cgk94,gk96,bgk97P}.
In the dissipative case, an essential ingredient is the time-separation
method introduced in \cite{ckv97P} in a one-dimensional context, and
afterwards exploited in Ref. \cite{cfk98}.
The point is that in developed turbulence there appears a natural
small parameter, \(1/P\!e\), where \(P\!e\) is the Peclet
number, measuring the relative strength of advection with respect
to diffusion. One would like to use
this parameter to develop an asymptotic theory for quantities
like the p.d.f. \({\cal P}(\varepsilon)\) of the dissipation field
\(\varepsilon\). However, such quantities are essentially non perturbative
in \(1/P\!e\). For instance, \(\varepsilon\) is exactly zero
without diffusion, but it has a non-zero limit as
\(1/P\!e\ne 0\), \(P\!e\rightarrow\infty\).
\footnote{It is worth noting that without dissipation
stationary limits for the p.d.f.'s
of the passive scalar field, of its gradients,
{\it etc.}, do not exist.
In the case of the direct cascade considered
in this paper an arbitrary small but finite dissipation constant $\kappa$
provides stationary distributions at large times
for all the physical quantities. However,
there are cases corresponding to the dynamical inverse cascade
when dissipation
does not lead to stationary p.d.f.'s for some observables \cite{ChKVl}.}

Following Refs. \cite{ckv97P,cfk98}, we show that an appropriate dynamical
formalism naturally introduces two time-scales, a short one
related to stretching, and a long one related to diffusion.
The time scale of the stretching fluctuations is of the order of the inverse
of the maximum Lyapunov exponent, while the whole time of stretching is
\(\ln P\!e\) times larger.

The paper is organised as follows:

In Sec. \ref{statement} we define the problem, recall the
statistical and kinematical concepts involved, introduce the
functional transformation which makes the problem solvable and
write down the corresponding functional Jacobian.

In Sec. \ref{timesep} we expose the time-separation method, showing
that the computation of \({\cal P}(\varepsilon)\) can be split in two
parts, which we will denote as small- and large-times averaging.
Suggestively, this procedure can be thought of as a two-step application of
a renormalisation group transformation.

In Secs. \ref{largetimes} and \ref{smalltimes} we perform
respectively large- and small-times averaging, reducing the computation of
the corresponding functional integrals to solvable auxiliary quantum-mechanical
problems.

In Sec. \ref{resulting} we put together all the pieces of the
computation and present the resulting p.d.f.  \({\cal P}(\varepsilon)\),
whose asymptotic behaviour is discussed as a function of
the space dimension \(d\).

\section{Statement of the problem}
\label{statement}
The advection of a passive scalar field $\theta(t,{\bf r})$ by the
incompressible smooth flow ${\bf v}(t,{\bf r})$ in $d$-dimensional
space is governed by the transport equation \begin{equation} (\partial_t
 +{\bf v}\cdot\nabla-\kappa\, \triangle) \,\theta=\phi\, ,\label{a1}\end{equation}
where $\phi(t,{\bf r})$ is an external source, $\kappa$ is the
diffusivity  and we assume a null initial condition at $t=-\infty$.
We assume that both ${\bf v}(t,{\bf r})$ and $\phi(t,{\bf
 r})$ are Gaussian independent random functions, $\delta$-correlated
in time \cite{kraichnan68,kraichnan74,ss94,cfkl95} so that, firstly:\begin{equation}
 \langle{ \phi( t_1, {\bf r}_1)\phi(t_2,{\bf r}_2)}\rangle= \delta(t_1-t_2)\chi(r_{12})\,,
 \label{a2}\end{equation} where $r_{12}\equiv|{\bf r}_1-{\bf r}_2|$,
the correlation $\chi(r_{12})$ decays on the scale $L$, and the constant
$P_2\equiv\chi(0)$ is the production rate of $\theta^2$. We consider
large Prandtl numbers, which correspond to a large viscosity-to-diffusivity
ratio, so that in the viscous interval \begin{equation} \langle v_\alpha(t_1,{\bf
 r}_1) v_\beta(t_2,{\bf r}_2)\rangle= \delta(t_1-t_2) \Bigl [V_0 \delta_{
 \alpha \beta} -{\frak D}(r^2\delta_{\alpha\beta}-r_\alpha r_\beta)-
 \frac{(d-1){\frak D}}{2} \delta_{\alpha\beta} r^{2}\Bigr]\,. \label{a6}
 \end{equation} Isotropy of the velocity statistics is here assumed.
The representation (\ref{a6}) is valid for scales smaller than the
velocity infrared cut-off $L_{\rm u}$ (which is supposed to be the
largest scale of the problem), since (\ref{a6}) represents the two
first terms of the expansion of the velocity correlation function
in powers of $r/L_{\rm u}$, and ${\frak D}\sim V_0/L_{\rm u}^2$. We
assume also that the inequality $P\!e^2\equiv {\frak D}L^2/2\kappa\gg1$
holds, guaranteeing the existence of a convective interval of scales
$r_{\rm diff}\ll r\ll L$ where the dominant effect to be taken into account
is the stretching of the velocity field. Here
$r_{\rm diff}=2\sqrt{\kappa/(d-1){\frak D}}$
is the mean diffusion length.

It follows from (\ref{a6}) that the correlation functions of the traceless
strain field, defined as $\sigma_{\alpha\beta}=\nabla_\beta v_\alpha$,
are ${\bf r}$-independent: \begin{equation} \langle\sigma_{\alpha\beta}(t_1)
 \sigma_{\mu\nu}(t_2)\rangle= {\frak D}\left[(d+1)\delta_{\alpha\mu}\delta_{\beta\nu}
 -\delta_{\alpha\nu}\delta_{\beta\mu} -\delta_{\alpha\beta}\delta_{\mu\nu}\right]
 \delta(t_1-t_2) \,. \label{ax6} \end{equation} This means that the
strain field $\sigma_{\alpha\beta}$ can be treated as a random function
of time $t$ only. To exploit this property it is convenient to pass
to a comoving reference frame, that is, to a coordinate frame whose
origin follows the motion of a Lagrangian particle of the fluid, so
that the velocity field can be approximated by
\( {\bf v} \simeq \hat\sigma (t){\bf r}  \),
with \( \langle\hat\sigma_{\alpha\beta}(t)\rangle=0 \)
\cite{cfkl95}. With this substitution,
(\ref{a1}) can be solved by Fourier transforming in the space coordinates
and integrating along characteristics:
\begin{equation}
\label{te1}
\theta _{{\bf k} }(t)=\int _{-\infty}^{t}\! {\rm d} t'\, \phi \big (t',\hat{{\cal W}} ^{\rm T} (t,t')
{\bf k} \big )\exp \Big [-\kappa \, {\bf k} \! \cdot \! \int _{t'}^{t}\! \! {\rm d} t_1
\hat{{\cal W}} (t,t_1)\hat{{\cal W}} ^{\rm T} (t,t_1){\bf k}\Big]
\end{equation}
where
\begin{equation}
\label{wop}
\hat{{\cal W}} (t,t')={\cal T}\exp
\left( \int _{t'}^{t}\hat\sigma (\tau ){\rm d} \tau \right)
\end{equation}
is the solution of the evolutive problem
\begin{equation}
\label{prt}
\dot {\hat{{\cal W}} }(t)=
\hat\sigma (t)\hat{{\cal W}} (t ),\;\quad \hat{{\cal W}} (t')={\bf 1}\, ,
\end{equation}
and \({\cal T}\) is the chronological ordering operator.
We will study stationary statistical properties of the passive scalar  at the fixed
time moment $t$. It is convenient to use
the symmetries of the measure of averaging over $\hat\sigma (\tau )$
with  respect to the transformations 
\(
\hat\sigma (\tau )\longrightarrow \hat\sigma (-\tau )
\)
and \(\hat\sigma(\tau)\to -\hat\sigma(\tau)\).
Changing now variables in the time integral in (\ref{te1}) 
we arrive to the expression:
\begin{equation}
\label{te2}
\theta _{{\bf k} }(t)=\int _{0}^{\infty}\! {\rm d} t'\, \phi 
\big (t-t',\hat{{\cal W}} ^{-1,{\rm T} }(t')
{\bf k} \big )\exp \Big [-\kappa \, {\bf k} \! \cdot \! \int _{0}^{t'}\! \! {\rm d} \tau
\hat{{\cal W}} ^{-1}(\tau )\hat{{\cal W}} ^{-1,{\rm T} }(\tau ){\bf k}\Big]
\, .
\end{equation}
Here $\hat{{\cal W}}(t)\equiv \hat{{\cal W}} (t,0)$ and invariance of the
random force $\phi $ averaging with respect to time inversion was used.
Let us now consider the statistics of the dissipation field
\[
\varepsilon =\kappa (\nabla \theta )^{2}\, .\]
The p.d.f.
of the \( \varepsilon  \) field can be written as
\begin{equation}
\label{pdf}
{\cal P}(\varepsilon )=\frac{1}{2\pi i}\int _{-i\infty }^{+i\infty }{\rm d} s\, e^{s\varepsilon }\, \int \frac{{\rm d} ^{d}m}{\pi ^{d/2}}\, e^{-m^{2}}\, \langle e^{-sQ}\rangle _{\hat\sigma }\, ,
\end{equation}
with
\begin{equation}
\label{H7}
Q=\frac{{\frak D}}
{P\!e ^{2}}\int _{0}^{\infty }{\rm d} t\,
\int \frac{{\rm d} ^{d}k}{(2\pi )^{d}}\,
\chi _{k}\,\,
\big ({\bf k} \! \cdot \! \hat{{\cal W}} (t){{\mathbf m}} \big )^{2}
\,\, \exp \left( -\frac{{\frak D}}{P\!e ^{2}}\,
{{\bf k} }\!\cdot\!{\hat\Lambda}(t){\bf k}\right)
\end{equation}
and
\begin{equation}
\label{ljam}
\hat\Lambda (t)=
\hat{{\cal W}} (t)\left( \int _{0}^{t}\hat{{\cal W}} ^{-1}(\tau )
\hat{{\cal W}} ^{-1,{\rm T} }(\tau ){\rm d} \tau \right) \hat{{\cal W}} ^{{\rm T} }(t)
\end{equation}

Here the integration over the auxiliary vector parameter \( {\mathbf m}  \) takes care
of combinatorics and summation over vector indices. The moments of
the dissipation field \( \varepsilon  \) can be recovered as
\begin{equation}
\label{nmoment}
\langle \varepsilon ^{n}\rangle =\int \frac{{\rm d} ^{d}m}{\pi ^{d/2}}\, e^{-m^{2}}\, \langle Q^{n}\rangle _{\hat\sigma }\,.
\end{equation}
The average \( \langle \ldots \rangle _{\hat\sigma } \) over the
traceless random strain field \( \hat\sigma  \) has to be performed with the 
probability
measure \( {\cal D}\hat\sigma (t)\exp (-S_{\hat\sigma }) \)
corresponding to (\ref{ax6}):
\begin{equation}
\label{S2}
S_{\hat\sigma }=\frac{1}{2d(d+2)\frak D}\int _{0}^{+\infty }\big [(d+1){\rm Tr}(\hat\sigma \hat\sigma ^{{\rm T} })+{\rm Tr}(\hat\sigma ^{2})\big ]{\rm d} t\,.
\end{equation}
The matrix \( \hat\Lambda (t) \) is invariant under right local
(time dependent) rotations \( \hat{{\cal W}} (t)\rightarrow \hat{{\cal W}} (t)\hat{{\cal R}}(t) \), and transforms covariantly under
left local rotations \( \hat{{\cal W}} (t)\rightarrow \hat{{\cal R}} (t)\hat{{\cal W}} (t) \). The quantity \( Q \) is invariant under left local
rotations, which for any fixed \( t \) can be absorbed in the \( {\rm d} ^{d}k \) integration.
The action \( S_{\hat\sigma } \) is invariant under both left and right
global rotations
and under the transformation \(\hat\sigma\to\hat\sigma^{\rm T}\).
Together with isotropy of the pumping function \( \chi  \),
invariance of \( S_{\hat\sigma } \) and \( \hat\Lambda  \) under right rotations
implies that \( Q\equiv Q({{\mathbf m}} ) \) is
a function of the square modulus \( z={{\mathbf m}} ^{2} \), so that we can substitute
\( {\mathbf m} \rightarrow \sqrt{z}{\bf n} _{0} \) in (\ref{H7}), with
\begin{equation}
\label{vem}
{\bf n} _{0}=\left( \begin{array}{c}
1\\
0\\
\vdots \\
0
\end{array}
\right) \, .
\end{equation}
The \( m \)-integration measure in (\ref{pdf}) and (\ref{nmoment}) can then
be substituted by
\begin{equation}
\label{osm}
\frac{1}{\Gamma (d/2)}z^{\frac{d}{2}-1}e^{-z}\,{\rm d} z
\end{equation}
After these substitutions \( Q \) maintains invariance under
left local rotations of \(\hat{\cal W}\).

Let us perform the Iwasawa decomposition of the evolution matrix \( \hat{{\cal W}}  \):
\begin{equation}
\label{decom}
\hat{{\cal W}} (t)=\hat{{\rm R}} \hat{{\rm D}} \hat{{\rm T}} ^{-1}\, ,
\end{equation}
where, for any fixed \( t \), \( \hat{{\rm R}}  \) is an \( {\rm SO}(d) \) rotation matrix, \( \hat{{\rm D}}  \) a diagonal matrix
with \( \det \hat{{\rm D}} =1 \), and \( \hat{{\rm T}}  \) an upper triangular matrix with \( 1\)'s on the diagonal.

The action \( S_{\hat\sigma } \) takes then the form: \begin{eqnarray}
\nonumber S_{\hat\sigma}
 &=& \frac 1{2{ d}{\frak D}} \int_0^\infty{\rm d} t\,\Bigl\{ {\rm Tr}\big(
 {\dot{{\hat{{\rm D}}}} {\hat{{\rm D}}}} ^{-1}\big) ^2+\frac 12{\rm Tr}\big[{ {\hat{{\rm D}}}}^2\big(
 { {\hat{{\rm T}}}} ^{-1} \dot{{ {\hat{{\rm T}}}}}\big) { {\hat{{\rm D}}}}^{-2}\big( {{\hat{{\rm T}}}}^{-1}\dot{
  { {\hat{{\rm T}}}}}\big) ^{\rm T}\big] \label{deo0}\\ &&\phantom{\frac 1{2{ d}{\frak D}}
 \int_0^\infty{\rm d} t\,}\quad-\frac{{ d}}{{ d}+2} {\rm Tr}\big[
 {\hat{{\rm R}}}  ^{-1} \dot{{\hat{{\rm R}}}}-\big({ {\hat{{\rm D}}}} { {\hat{{\rm T}}}} ^{-1}\dot{{ {\hat{{\rm T}}}
  }} { {\hat{{\rm D}}}}^{-1}\big)_{\rm a}\big] ^2\Bigr\}\,. \label{deo} \end{eqnarray}
 where $\left( \ldots \right) _{\rm a}$ denotes the antisymmetric
part.

The invariance of \( Q \) under local rotations of \(\hat{\cal W}\)
implies that \(Q\) does not depend on \(\hat {\rm R}\), so that
the \( \hat{{\rm R}}  \) variables
can be integrated out from the very beginning.
The resulting effective action is equal to \( S_{\hat\sigma } \) with the last trace-term
dropped. It is then convenient to parametrize
\begin{equation}
\label{diae}
\hat{{\rm D}} =\left( \begin{array}{cccc}
e^{\rho _{1}}, & 0, & \ldots , & 0\\
0, & e^{\rho _{1}}, & \ldots , & 0\\
\ldots , & \ldots , & \ldots , & \ldots \\
0, & 0, & \ldots , & e^{-(\rho _{1}+\cdots +\rho _{d-1})}
\end{array}
\right)
\end{equation}
and
\begin{equation}
\label{triae}
\hat{{\rm T}} =\left( \begin{array}{ccccc}
1, & \eta _{1}, & \eta _{2}, & \ldots , & \eta _{d-1}\\
0, & 1, & X_{23}, & \ldots , & X_{2,d}\\
\ldots , & \ldots , & \ldots , & \ldots , & \ldots \\
0, & 0, & 0, & \ldots , & 1
\end{array}
\right) \, ,
\end{equation}
denoting
by \( \eta _{j}\equiv X_{1,j+1} \)
the elements of the first row of \( \hat{{\rm T}}  \), which, together with the
\( \rho _{j} \), will be seen to be the only relevant dynamical variables. In order
to have more compact expressions it will
be also convenient in what follows to denote the sum \( \rho _{1}+\cdots +\rho _{d-1} \) by \( -\rho _{d} \) and to set \(X_{jj}\equiv 1\).

The initial condition \( \hat{{\cal W}} (0)= {\bf 1} \) implies
\begin{eqnarray}
\label{inc0} \rho _{j}(0) & = & 0,\quad \eta _{j}(0)\, =\, 0,\quad j=1,\ldots ,d-1\, ,\\
\label{inc} X_{mn}(0) & = & 0,\quad 2\le m<n\le d\, .
\end{eqnarray}
With the substitution \( \hat\sigma \rightarrow \big (\hat{{\rm R}} ,\hat{{\rm D}} ,\hat{{\rm T}} \big ) \) we introduced
in the functional integral a set of new ``collective'' coordinates,
by means of a non-linear, non-local variable transformation. The Jacobian
of the transformation is given by
\begin{equation}
\label{jak}
{\cal J}={\cal J}_{{\rm ul}}\cdot {\cal J}_{{\rm l}}=\prod _{t=0}^{\infty }\exp \left[ 2\sum _{j=1}^{d-1}(d-j)\rho _{j}(t)\right] \cdot \exp \left[ \int _{0}^{\infty }\sum _{j=1}^{d-1}(d-j)\dot\rho _{j}(\tau ){\rm d} \tau \right] \, .
\end{equation}
The details of the derivation are given in the Appendix.
The local part \( {\cal J}_{{\rm l}} \) of the Jacobian coincides with the one computed in
Ref. \cite{cgk94}.

With the decomposition (\ref{decom}), and after the integration of the \( \hat{{\rm R}}  \) variables,
the probability measure takes the form
\[
\prod _{j=1}^{d-1}{\cal D}\rho _{j}{\cal D}
\eta _{j}\!\!\prod _{2\le m<n\le d}{\cal D}X_{mn}\: {\cal J}(\rho )\, \exp \left( -S_{{\rm eff}}[\rho ,\eta ,X]\right) \, ,\]
with
\begin{eqnarray}
\label{azione2} S_{{\rm eff}} & = & \frac{1}{2d{\frak D}}\int _{0}^{\infty }{\rm d} t\Big [\sum _{j=1}^{d}\dot\rho _{j}^{2}+\frac{1}{2}\sum _{1\le i<j\le d}e^{2(\rho _{i}-\rho _{j})}\xi _{ij}^{2}\Big ]
\end{eqnarray}
and \( \hat\xi =\hat{{\rm T}} ^{-1}\dot {\hat{{\rm T}} } \). Let us also rewrite
(\ref{H7}) as
\begin{equation}
\label{qun}
Q=\frac{z{\frak D}}{P\!e ^{2}}\int _{0}^{\infty }{\rm d} t\int \frac{{\rm d} ^{d}k}{(2\pi )^{d}}\, \chi _{k}\, k_{1}^{2}\, e^{2\rho _{1}}\, \exp \left( -\frac{{\frak D}}{P\!e ^{2}}\, {\bf k} \! \cdot \! \hat\Lambda (t){\bf k} \right) \, ,
\end{equation}
whereby fixing the residual left-rotation symmetry.

\section{Time separation}
\label{timesep}
We shall now expose a method that allows to compute \( {\cal P}(\varepsilon ) \) in the limit
of large Peclet numbers, \em i.e.\em , in a regime characterized by
a large ratio of advection to diffusion at the pumping scale.

The moments of the dissipation field \( \varepsilon  \) have finite,
non zero limits as diffusivity
tends to zero. Thus, \({\cal P}(\varepsilon)\) is a non-perturbative
quantity in the small parameter \( 1/P\!e  \).

The time separation method, which was introduced in Ref. \cite{ckv97P} in a
one-dimensional context and generalized to \(d=2\) in \cite{cfk98},
is the proper tool for performing such a
non-perturbative calculation. As a matter of fact, the dynamical formalism
itself  reveals the presence of two different time scales:
a short one, related to diffusion, and a long one, related to advective
stretching. Taking properly into account the two time scales it is
possible to exploit non-trivially the presence of the large parameter
\( P\!e  \) and to develop an asymptotic theory that captures the dominant term
in \( {\cal P}(\varepsilon ) \) with respect to \( P\!e  \).

The starting point is that, as is clear from the structure of
(\ref{qun}),
the main contribution to the moments
\(Q^n\) is obtained in the limit \(P\!e\to\infty\) for
\(\rho_1\sim \ln P\!e\).
On the other hand, one has the following result (see \cite{gk96})
for the hierarchy of the Lyapunov exponents:
\begin{equation}
\label{lyapunov}
\langle \rho _{j}(t)\rangle _{\hat\sigma }=\frac{d}{2}{\frak D}(d-2j+1)t
\equiv \bar\lambda _{j}t\, ,
\end{equation}
showing that the relevant values of \(\rho_1\) are realized for times
\begin{equation}t\simeq \frac{2}{d(d-1){\frak D}}\ln P\!e \, .
\label{tempo}
\end{equation}
For large times the elements of the matrix \(\Lambda(t)\) can
be estimated roughly as
\[
\Lambda _{jl}(t)\sim \delta _{jl}e^{2\rho _{j}(t)},\quad t\rightarrow \infty \]
and from (\ref{lyapunov}) it follows that at times of the order (\ref{tempo})
the components \(\Lambda_{jl}(t)\) with \(j,l\ne 1\) are small compared
with \(P\!e^2\) and can be neglected in the diffusive exponent (\ref{qun}).
Thus, only \(\Lambda_{11}\) effectively survives.

Let us then introduce a separation time \( t_{0} \) such that \( 1\ll d^{2}{\frak D}\, t_{0}\ll \ln P\!e  \).
In the limit \( P\!e \rightarrow \infty  \) this separation time will disappear from the final
answer. The contribution to \( Q \) from the interval \( 0<t<t_{0} \) is parametrically
small, so
that we can substitute \( 0\rightarrow t_{0} \) as a lower integration limit in (\ref{qun}).

Observe now that for the trajectories
contributing to \(\langle Q^n\rangle\)
the \( \xi _{ij} \) fields, being multiplied
in the action (\ref{azione2}) by a factor \( \exp (2\rho _{i}-2\rho _{j})\gg 1 \),
will be exponentially depressed for
\( t\gg 1 \). The components of \( \hat{{\rm T}}  \) will therefore be frozen to constant values
for those times. This is in particular true for the components \( \eta _{j} \) of
the first row.

Taking into account the freezing of \( \hat{{\rm T}}  \) and
writing explicitly \( \hat\Lambda  \) in
the parame\-tri\-za\-tion (\ref{diae},\ref{triae}) it is seen that the integral in (\ref{ljam}) saturates
for \( t\simeq t_{0} \), so that the \em upper \em integration limit in (\ref{ljam}) can be substituted
with \( t_{0} \). As a final result, in the limit \( P\!e \rightarrow \infty  \), (\ref{qun})
asymptotically reduces
to
\begin{eqnarray}
\label{qup}  &  & \quad Q\, =\, \frac{z{\frak D}}{P\!e ^{2}}\int _{t_{0}}^{\infty }{\rm d} t\int \frac{{\rm d} ^{d}k}{(2\pi )^{d}}\, \chi _{k}\, k_{1}^{2}\, e^{2\rho _{1}(t)}\, \times \\
\label{qup0}  &  &
\times \exp \left[ -\frac{\frak D}{P\!e ^{2}}\, k_{1}^{2}e^{2\rho _{1}(t)}\, 
{\bf n} _{0}\!
\cdot \! \int _{0}^{t_{0}}{\rm d} \tau \hat{{\rm T}} ^{-1}(t_{0})
\hat{{\cal W}} ^{-1}(\tau )\hat{{\cal W}} ^{-1,{\rm T} }(\tau )
\hat{{\rm T}} ^{\rm T}(t_0){\bf n}_0\right] \nonumber
\end{eqnarray}
Observe that all the dependence of \( Q \) on the stochastic variables
labelled by small times \( t<t_{0} \) is now concentrated in the square-bracketed
exponent of (\ref{qup}). The action (\ref{azione2}) can also be written as a sum
\( S_{{\rm eff}}^{<}+S_{{\rm eff}}^{>} \) of
a small- and a large-times part. This makes clear the statement that
integration over small-times variables, which describe a mainly dissipative
process, can be separated from integration over large-times (\( t>t_{0} \)) variables,
which describe multiplicative stretching. Time separation appears
therefore as a natural consequence of the dynamic formalism.

The small-times and large-times variables are coupled only by the
condition of continuity at \( t=t_{0} \):
\begin{equation}
\label{bcond}
\rho _{j}^{>}(t_{0})=\rho _{j}^{<}(t_{0}),\quad \eta _{j}^{>}(t_{0})=\eta _{j}^{<}(t_{0}),\quad X^{>}_{mn}(t_{0})=X_{mn}^{<}(t_{0})\, .
\end{equation}
In order to fully implement time separation
we perform the reparametrization
\[
\hat{{\cal W}} (t)\rightarrow \hat{{\cal W}} (t)\hat{{\rm T}} ^{-1}(t_{0})\, .\]
which implies the substitution
\( \hat{{\rm T}} (t_{0})\rightarrow {\bf 1} \) in (\ref{qup}) and modifies
the boundary conditions (\ref{inc0},\ref{inc}) and (\ref{bcond}) as follows:
\begin{equation}
\label{insmall}
\rho _{j}^{<}(0)=0,\quad \eta ^{<}_{j}(t_{0})=X_{mn}^{<}(t_{0})=0\, ,
\end{equation}
for the small-times
fields, and
\begin{equation}
\label{inlarge}
\rho _{j}^{>}(t_{0})=\rho ^{<}_{j}(t_{0}),\quad \eta _{j}^{>}(t_{0})=X^{>}_{mn}(t_{0})=0
\end{equation}
for the large times.

Thus, all the dissipative effects are encoded in the p.d.f.
of the only variable
\begin{equation}
\label{obna}
\mu ={\bf n} _{0}\cdot \int _{0}^{t_{0}}{\rm d} \tau \, \hat{{\cal W}} ^{-1}(\tau )\hat{{\cal W}} ^{-1,{\rm T} }(\tau ){\bf n} _{0}\, ,
\end{equation}
which appears as a parameter in the large-times averaging. In order to
compute the total generating function
\begin{equation}
\label{prf}
{\cal P}_{s,z}=\langle e^{-sQ}\rangle
\end{equation}
we shall integrate \em first
\em on the large-times variables (essentially, only \( \rho _{1} \)), find a p.d.f.
\( {\cal P}^{>}_{s,z} \) which depends on \( \mu  \) as a parameter, and then complete the computation
by integrating over the small-times p.d.f. \( {\cal P}^{<}(\mu ) \) This procedure can be
suggestively thought as a renormalization group transformation consisting
of only two steps.

Let us now outline the first step. First of all, the variables \( \eta _{j}^{>} \),
\( X_{mn}^{>} \) can be integrated out completely, since they do not appear in (\ref{qup}).
Of the
\( \rho ^{>}_{j} \) variables, the only \( \rho _{1}^{>} \) appears in (\ref{qup}). Integration over \( \rho _{1}^{>} \) can be
separated from the trivial integration over \( \rho _{2}^{>},\ldots ,\rho _{d-1}^{>} \) by the simple shift
\begin{equation}
\label{roi}
\rho _{1}^{>}(t)=\tilde\rho _{1}(t)+\rho _{1}(t_{0}),\quad \rho _{j}^{>}(t)=\tilde\rho _{j}(t_{0})-\frac{\tilde\rho _{1}(t)}{d-1}+\rho _{j}(t_{0}),\; j=2,\ldots ,d-1\, ,
\end{equation}
which
also substitutes the initial condition \( \rho _{j}^{>}(t_{0})=\rho _{j}^{<}(t_{0}) \) with the simpler
\begin{equation}
\label{simpler}
\tilde\rho _{j}(t_{0})=0,\quad j=1,\ldots ,d-1\, .
\end{equation}
Finally,
we are left with an integration over the effective measure already
obtained in Ref. \cite{cgk94}:
\begin{equation}
\label{mef}
{\cal D}\rho _{1}\, \exp \left[ -\frac{1}{2(d-1){\frak D}}\int _{t_{0}}^{\infty }{\rm d} t\, \left( \dot {\tilde\rho _{1}}-\frac{d(d-1){\frak D}}{2}\right) ^{2}\right]
\end{equation}
and with the factor
\begin{equation}
\label{jano}
{\cal J}_{0}(\rho )=\exp \left[ \sum _{j=1}^{d-1}(d-j)\rho _{j}(t_{0})\right] \, ,
\end{equation}
which will
be part of the small-times averaging measure.

\section{Large-times averaging}
\label{largetimes}
Let us now apply the previous considerations to the actual evaluation
of \( {\cal P}^{>}_{s,z} \), defined as the average of \( e^{-sQ} \)
with respect to the measure (\ref{mef})
with the constraint \( \tilde\rho _{1}(t_{0})=0 \). The evaluation of the corresponding path integral
can be reduced via the Feynman-Kac formula to the solution of the
following auxiliary quantum mechanical problem in the space of the
variable \( \rho \equiv \tilde\rho _{1} \):
\begin{eqnarray}
\label{psz} {\cal P}^{>}_{s,z} & = & 
\lim _{T\rightarrow \infty }e^{-\frac{d^{2}(d-1){\frak D}}{8}(T-t_{0})}\,\, 
\langle \delta (\rho )|\exp [-(T-t_{0})\hat{{\cal H}} ^{>}]|e^{\frac{d}{2}\rho }\rangle \nonumber \\
\label{qmp}  & = & \lim _{T\rightarrow \infty }
e^{-\frac{d^{2}(d-1){\frak D}}{8}(T-t_{0})}\,\, \Psi (T-t_{0};\rho =0)\, ,
\end{eqnarray}
with
\begin{eqnarray}
\label{accapiu} \hat{{\cal H}} ^{>} & = & -\frac{(d-1){\frak D}}{2}\,
\frac{\partial ^{2}}{\partial\rho ^2}+V^{>}(\rho )\, ,\\
\label{vpiu} V^{>}(\rho ) & = & sz\beta \frac{{\frak D}}{P\!e ^{2}}\cdot
e^{2\rho}\cdot
\Xi \left( \sqrt{\frac{{\frak D}}{P\!e ^{2}}\beta \mu }\, e^{\rho }\right) ,\quad
\beta =e^{2\rho (t_{0})}\, ,\\
\label{csi} \Xi (y) & = &
\int \frac{{\rm d}^d k}{(2\pi)^d}\,\chi _{k}\, k_1^2 e^{-k_1^{2}y^{2}}\, .
\end{eqnarray}
In (\ref{psz}), the ``wave function'' \( \Psi  \) is the solution of
the initial value problem
\begin{eqnarray}
\label{auxquant} \hat{{\cal H}} ^{>}\Psi  & = & -\partial _{t}\Psi \, ,\\
\label{auxbound} \Psi (t=0;\rho ) & = & e^{\frac{d}{2}\rho }\, .
\end{eqnarray}
For \( t\rightarrow \infty  \) the behaviour of \( \Psi (t,\rho ) \) is dominated
by the lowest eigenvalue of the corresponding stationary problem,
which can be easily determined since the potential \( V^{>}(\rho ) \) vanishes at \( \rho \rightarrow \infty  \),
leaving us with a free Hamiltonian:
\[
\hat{{\cal H}} ^{>}e^{\frac{d}{2}\rho }\simeq 
-\frac{(d-1){\frak D}}{2}\frac{d^{2}}{4}\, 
e^{\frac{d}{2}\rho },\quad \rho \rightarrow \infty \, .\]
So, for \( t\rightarrow \infty  \):
\begin{equation}
\label{psiasympt}
\Psi (t;\rho )\simeq e^{\frac{d^{2}(d-1){\frak D}}{8}\, t}\, e^{\frac{d}{2}\rho }\, \Phi (y)\, ,
\end{equation}
with \( \Phi (y) \) the solution
of the following boundary problem in the variable 
\( y=\sqrt{{\frak D}\beta \mu /P\!e ^{2}}\; e^{\rho } \):
\begin{eqnarray}
\label{lteq}
\label{dlinur} \left[ -\frac{1}{y^{d+1}} 
\frac{\partial}{\partial y} \, y^{d+1}
\frac{\partial}{\partial y}+
\frac{2sz}{(d-1){\frak D}\mu }\, \Xi (y)\right] \, \Phi (y) & = & 0\, ,\\
\label{codl} \Phi (y)\rightarrow 1,\quad y\rightarrow \infty ;\qquad y\Phi (y)\rightarrow 0,\quad y\rightarrow 0\, .\; \phantom {\Bigg [..} &  &
\end{eqnarray}
Note that \( \Phi (y) \) can
depend on the parameters \( s,\, z,\, {\frak D} \) and \( \mu  \) only 
through their combination
\begin{equation}
\label{lambdaduesz}
\lambda =\frac{2sz}{(d-1){\frak D}\mu }\, ,
\end{equation}
and
does not depend on \( P\!e  \).

Finally, from (\ref{qmp},\ref{psiasympt}) it follows\em
\begin{equation}
\label{ppsi}
{\cal P}^{>}_{s,z}=\Phi \left( \sqrt{\frac{{\frak D}}{P\!e ^{2}}\beta \mu }\right) \, .
\end{equation}
\em Here \( {\cal P}^{>}_{s,z} \) depends on \( P\!e  \) only through
the argument of \( \Phi  \). In the limit \( P\!e \rightarrow \infty  \) one simply finds
\begin{equation}
\label{ppsin}
{\cal P}^{>}_{s,z}=\Phi (0)
\end{equation}
For a generic given \( \chi _{k} \), one can solve (\ref{dlinur}) iteratively in terms of the
small potential \( s\Xi (y) \), \( s\rightarrow 0 \), finding explicit expressions for all the ``large-times''
moments of \( Q \):
\begin{eqnarray}
\label{momenti} {\cal P}^{>}_{s,z} & = & 1+\sum _{k=1}^{\infty }(2k-1)!!\cdot \left( \frac{2s\beta }{(d-1){\frak D}\mu }\right) ^{k}\times \\
 &  & \times \int _{\scriptsize {\begin{array}{c}
y_{2k}<y_{2k-1}\\
y_{2k}<y_{2k+1}
\end{array}
}}{\rm d} y_{1}\cdots {\rm d} y_{2k}\, \prod _{l=1}^{k}\left( \frac{y_{2l}}{y_{2l-1}}\right) ^{d+1}\Xi (y_{2l})\nonumber
\end{eqnarray}
The moments are here expressed directly in terms of the integral transform
(\ref{csi}) of the pumping function \( \chi _{k} \).


\section{Small times averaging}
\label{smalltimes}
In this section we shall compute
\begin{equation}
\label{lapma}
{\cal P}_{s}^{<}=\langle e^{-s\mu }\rangle \, .
\end{equation}
The explicit expression of \( \mu  \) is
\[
\mu =\int _{0}^{t_{0}}V^{<}{\rm d} t\]
with
\begin{equation}
\label{potma}
V\equiv V^{<}=e^{-2\rho _{1}}+\eta _{1}^{2}e^{-2\rho _{2}}+\eta _{2}^{2}e^{-2\rho _{3}}\cdots +\eta ^{2}_{d-1}e^{-2\rho _{d}}
\end{equation}
In
what follows we will omit the \( (\ldots )^{<} \) index on the \( \rho _{j} \) and \( \eta _{j} \) variables. Along
the lines previously exposed in the case of large-times averaging,
the computation of \( {\cal P}^{<}_{s} \) is again reduced to the solution of an auxiliary
quantum mechanical problem:
\begin{eqnarray}
\label{smaq}  &  & \qquad \quad {\cal P}_{s}^{<}=\exp \left( -\frac{d^{2}(d^{2}-1){\frak D}}{24}\, t_{0}\right) \times \\
 &  & \times \Big \langle \prod _{j=1}^{d-1}e^{(d-j)\rho _{j}}\, \delta (\eta _{j})\prod _{2\le m<n\le d}\delta (X_{mn})\Big |\exp \big (-t_{0}\hat{{\cal H}} ^{<}\big )\Big |\prod _{j=1}^{d-1}\delta (\rho _{j})\Big \rangle \, .\nonumber
\end{eqnarray}
Here
\[
\hat{{\cal H}} ^{<}=\hat{{\cal H}} _{0}+sV\, ,\]
and
\( \hat{{\cal H}} _{0} \) is the quantum Hamiltonian corresponding to the classical action
(\ref{azione2}):
\begin{eqnarray}
\hat{\cal H}_0&=&-\frac{(d-1){\frak D}}{2}\sum_{j=1}^{d-1}\frac{\partial^2}{\partial{\rho_j}^2}
+{\frak D}\sum_{0<i<j<d}\frac{\partial^2}{\partial{\rho_i\partial\rho_j}} \\
&&-d{\frak D} \sum_{1\le j<i\le d}e^{2 \rho_i-2\rho_j}\Big(\sum_{k\le j; \ k<i}X_{kj}
\frac{\partial}{\partial{X_{ki}}}\Big)^2
\label{explham}
\end{eqnarray}
Due to the the triangular structure of the \( X_{mn} \) and \( \eta _{j} \) variables
there arise no ordering problems.

The seemingly intractable quantum mechanical problem in \( \frac{1}{2}d(d+1)-1 \) dimensions
can in fact be reduced to a solvable, one-dimensional problem thanks
to the presence of \( \frac{1}{2}d(d+1)-2 \) symmetries.
First of all, the initial wave function in (\ref{smaq}) does not depend on the
\(X_{jl}\) variables. The evolution operator \(\exp(-t_0\hat{\cal H}^<)\)
introduces a dependence on the \(\eta_j\) variables, but not on the remaining
\(X_{ij}\) (\(i>1\)), as is clear from the explicit expressions (\ref{potma})
and (\ref{explham}).
We are thus led to consider the reduction \(\hat{\cal H}^{(\rho,\eta)}\) of \(\hat{\cal H}^<\)
on the space of functions depending only on \(\rho_j\) and \(\eta_j\):
\begin{eqnarray}
\hat{{\cal H}}^{(\rho,\eta)}  & = & -\frac{(d-1){\frak D}}{2}
\sum _{j=1}^{d-1}\frac{\partial^2}{\partial\rho _{j}^2}+
{\frak D}\!\!
\sum _{0<i<j<d}\frac{\partial^2}{\partial\rho _{i}\partial\rho _{j}} \\
 &  & -d{\frak D}\sum _{j=2}^{d}\Big (e^{-2(\rho _{1}-\rho _{j})}+
\sum _{k=2}^{j-1}e^{-2(\rho _{k}-\rho _{j})}\eta _{k-1}^{2}\Big )
\frac{\partial^2}{\partial\eta _{j-1}^2} + V\, .
\end{eqnarray}
We can then forget altogether about the \( X_{jl} \) variables and just compute
\begin{equation}
\label{wavef}
\Big \langle e^{\sum_{j=1}^{d-1}(d-j)\rho _{j}}\delta (\eta _{1})\cdots \delta (\eta _{d-1})
\Big |\exp \big (-t_{0}\hat{{\cal H}}^{(\rho,\eta)}\big )\Big |
\delta (\rho _{1})\cdots \delta (\rho _{d-1})\Big \rangle \, .
\end{equation}
The reduced Hamiltonian \(\hat{\cal H}^{(\rho,\eta)}\) depends on \(2d-2\) variables.
Let us use now the fact that the action \( S^{<}=S^{<}_{{\rm eff}}+s\mu  \)
is invariant under the
global left transformations
\begin{equation}
\label{tratt}
\hat{{\rm T}} (t)\rightarrow \hat\Theta \hat{{\rm T}} (t)\, ,
\end{equation}
where we take \( \hat\Theta = {\bf 1}+\delta \hat\Theta  \),
with \( \delta \hat\Theta  \) having non zero values only in the first column
and on the diagonal, and satisfying \( {\rm Tr}(\delta \hat\Theta )=0 \)
and \( (\delta \hat\Theta )_{00}=0 \).
This way we generate
\( 2d-3 \) symmetries of \( S^{<} \), which are readily identified with rotations in
the space of the variables \( \eta _{j-1}e^{2\rho _{j}} \)
and shifts of the \( \rho _{j} \) accompanied by corresponding
rescalings of the \( \eta _{j} \).
The corresponding variations of \( \hat{{\rm D}}  \) and \( \hat{{\rm T}}  \)
can be read out from \begin{equation} \left( {\rm \hat{D}\hat{T}}^{-1}\delta
 \hat{\Theta}{\rm \hat{T}\hat{D}}  ^{-1}\right) _{\rm s}=\left(
 \delta {\rm \hat{D}\hat{D}}^{-1}-{\rm \hat{D}\hat{T}  }^{-1}\delta
 {\rm \hat{T}\hat{D}}^{-1}\right) _{\rm s} \label{varii} \end{equation}
where $\left( \ldots \right) _{\rm s}$ denotes the symmetric part.
The corresponding integrals of the motion are also easily found.
They locally generate \(2d-3\) ``angular" directions, transverse to the
``radial" direction labelled by \(V\).
We will not need their explicit expression: it will be enough to
determine the ``radial" part
of \( \hat{{\cal H}} ^{(\rho ,\eta )} \) by the transformation
\( \partial _{x}=\partial V/\partial x\cdot \partial _{V} \) \( +(\hbox{radial part}) \),
with \( x \) substituted by \( \rho _{j} \),
\( \eta _{j} \).
In the \(t_0\gg1\) limit (\ref{wavef}) can be
computed explicitly as
\begin{equation}
\label{psless}
{\cal P}^{<}_{s}=\int _{-\infty }^{\infty }{\rm d} \eta _{1}\cdots
\int _{-\infty }^{\infty }{\rm d} \eta _{d-1}\, \Phi (\rho _{j}=0,\eta _{j})\, ,
\end{equation}
with \( \Phi  \) the solution of the stationary problem
\begin{eqnarray}
\label{lastham}  &  & \hat{{\cal H}} ^{(\rho ,\eta )}\Phi (\rho ,\eta )\, =\, 0\, ,\\
\label{lastbound1}  &  & \Phi (\rho ,\eta )\, \simeq \, \delta (\eta _{1})\cdots \delta (\eta _{d-1})\, \exp \Big [\sum _{j=1}^{d-1}(d-j)\rho _{j}\Big ]\, ,\quad \rho \rightarrow \infty \, ,\\
\label{lastbound2}  &  & \Phi (\rho ,\eta )\, \simeq \, 0\, ,\quad \rho \rightarrow -\infty \, .
\end{eqnarray}
Let us consider a maximally symmetric solution \( \Phi  \)
having the form
\begin{equation}
\label{phirhoeta}
\Phi (\rho ,\eta )=\exp \Big [\sum _{j=1}^{d-1}(d-j)\rho _{j}\Big ]{\cal F}(V)\, ,
\end{equation}
with \( {\cal F}(V) \) satisfying the ``radial" equation
\begin{equation}
\label{3dur}
2(d-1)V^{2}{\cal F}''+(d-1)(d+2)V{\cal F}'-\frac{sV}{{\frak D}}{\cal F}=0\, .
\end{equation}
A solution of (\ref{3dur}) is:
\begin{equation}
{\cal F}(V)=V^{-d/4}K_{d/2}
\left(\sqrt{\frac{2sV}{(d-1){\frak D}}}\right)\, ,\end{equation}
where \( K_{d/2} \) is the
Macdonald function of order \( d/2 \).
The properties of
the Macdonald functions \cite{gr94} allow to verify directly that
(\ref{phirhoeta}) satisfies the boundary 
conditions (\ref{lastbound1},\ref{lastbound2}).
This also shows \em a posteriori \em that
the boundary conditions (\ref{lastbound1},\ref{lastbound2}) are
asymptotically invariant under the symmetries of the problem.

The integration (\ref{psless}) can be explicitely performed, giving
\begin{equation}
{\cal P}_{s}^{<}={\rm const}\cdot \exp 
\left(-\sqrt{\frac{2s}{(d-1){\frak D}}}\right)\end{equation}
whose Laplace transform gives the p.d.f. of
\( \mu  \):
\begin{equation}
\label{rama}
{\cal P}^{<}(\mu )=\frac{1}{\sqrt{2\pi (d-1){\frak D}}\: 
\mu ^{3/2}}\, \exp \left( -\frac{1}{2(d-1){\frak D}\mu }\right) \, .
\end{equation}
One can check the soundness of the whole computation by verifying
on the first moment of \( \varepsilon  \) that the constraint of energy conservation
\[
\langle \varepsilon \rangle =\frac{1}{2}\int \frac{{\rm d} ^{d}k}{(2\pi )^{d}}\, \chi _{k}\]
is
satisfied.

\section{Resulting distribution function}
\label{resulting}
In order to restore the total p.d.f. ${\cal P}\left( \varepsilon \right)
 $ we should take the function ${\cal P}^{>}_{s,z} $ expressed in (\ref{ppsin})
through the solution of (\ref{dlinur}), then average over $\mu $ with
the weight (\ref{rama}) and over $z$ with the measure (\ref {osm}),
and finally perform an inverse Laplace transform. As already noticed,
$\Phi \left( y=0\right) $ depends
on the parameters $s,\,z,\,{\frak D}$ and $\mu $ only through 
their combination $\lambda$ (see (\ref{lambdaduesz})): \begin{equation}
 \Phi \left( 0\right) =f \left( \lambda \right)\, \label{psil} \end{equation}
 The resulting ${\cal P}\left( \varepsilon \right) $ can be expressed
via $f \left( \lambda \right) $ as:
\begin{equation}
{\cal P}
\left( \varepsilon \right) =
\frac {\sqrt{\frak D}}
{2\pi \Gamma \left( d/2\right)\ \sqrt\varepsilon}
\int\limits_{-\infty }^{+\infty }
{\rm d} x
\int\limits_0^{+\infty}
{\rm d} z\,\,
f \left( ix\right)
\left( -ix\right)^{  \frac d2-1}z^{d-2}
\exp
\left( ixz^2-
\frac{\sqrt{\varepsilon {\frak D}}}z\right)
\label{rezu}
\end{equation} The function $f \left( \lambda \right) $ has poles
on the real negative semiaxis; the asymptotics of ${\cal P}\left(
 \varepsilon \right) $ is defined by the pole closest to the origin, 
which we denote
as $-\lambda ^{*}$ ($\lambda^{*}>0$): \begin{equation} {\cal P}\left(
 \varepsilon \right) \sim \exp \left[ -\frac{3}{2}
 \left(2 \lambda ^{*}{\frak D}\varepsilon \right) ^{1/3}\right] ,
\,\,\varepsilon
 \rightarrow \infty \, .\label{asyre} \end{equation}
The point is that when \(\lambda\) approaches the value \(-\lambda^*\)
the Hamiltonian \(\hat{\cal H}^>\) develops 
a ground state energy
\[ E_0<-\frac{d^2(d-1){\frak D}}{8}\, \]
and \(f(\lambda)={\cal P}_{s,z}^>\) given by (\ref{psz}) becomes infinite.
We therefore have
\[\lambda^*=\frac{d^2}{4 \xi_*}\, \]
with \(\xi_*\) the maximum of the function \(y^2\Xi(y)\)
({\em cfr.} (\ref{accapiu},\ref{vpiu})).
This finally gives
\begin{equation} {\cal P}\left( \varepsilon \right) \sim
 \exp \left[ -{\rm const}\left( d^2{\frak D}  \varepsilon \right)
 ^{1/3}\right]\,  ,\,\,\ 
\varepsilon \rightarrow \infty \, .\label{bdas} \end{equation}

\section{Discussion}
\label{conclusions}
An important point to be noticed in the previous computation,
expecially with regard to the asymptotic expression (\ref{bdas}),
is that we consider in this paper \(P\!e\) as the largest
number in the theory.
For any large, but finite \(P\!e\) this means that
the asymptotics (\ref{bdas}) is valid for
\(\varepsilon/\varepsilon_0\gg 1\)
(where \(\varepsilon_0=\langle\varepsilon\rangle\))
but smaller than \(\ln P\!e\).
There is here a crucial difference with the computation of the p.d.f. \(P(\theta)\)
of the passive scalar \(\theta\) itself, which was discussed in the
papers \cite{cfkl95,cgk94,bgk97P,kls98P}.
If \(\theta\ll\ln P\!e\) the p.d.f. \(P(\theta)\)
has a Gaussian form and can be obtained by taking into account
only the variable \(\rho_1\) corresponding to the highest
Lyapunov exponent.
However, if \(\theta\gtrsim\ln P\!e\), this approach does not
give the correct exponent
for \(d>2\), as it was observed in \cite{bgk97P}.
The correct asymptotic of \(P(\theta)\) for \(\theta\gg\ln P\!e\) was
computed in \cite{bgk97P}; the complete p.d.f was then
restored in \cite{kls98P}, up to logarithmic corrections.

Let us estimate the domain of validity of our expressions
(\ref{rezu},\ref{bdas}), which were obtained in the limit
\(P\!e\to\infty\).
In the computation,
when \(\rho_1\sim\ln P\!e\) we neglected the dependence of \(Q\)
on \(\rho_2,\ldots,\rho_{d-1}\).
However,
due to the peculiar dependence of the potential on \(\rho_1\),
there exist trajectories for which \(\rho_1\) remains
confined in a finite domain around \(\ln P\!e\),
while the variables \(\rho_j\), \(j=2,\ldots,d-1\)
freely evolve, eventually reaching values of the order \(\ln P\!e\).
This happens for times of the order \(t^*\sim \ln P\!e\).
The corresponding value of \(\varepsilon\)
is \(\varepsilon_{\rm lim}\sim \varepsilon_0
t^*\sim\varepsilon_0\ln P\!e\).
For \(\varepsilon\gtrsim \varepsilon_{\rm lim}\)
the behaviour of \({\cal P}(\varepsilon)\)
is probably modified.

The exponent of the p.d.f.'s tail (\ref{bdas}) up to $d$-dependent
numerical coefficient coincides with the one-dimensional results
\cite{ckv97P}. This means that the configurations of the passive
scalar field contributing into (\ref{bdas})  are effectively one-dimensional,
as it was noted in \cite{BalF}.

\begin{center}{\bf Acknowledgements }\end{center}

We are grateful to
E.~Balkovsky, M.~Chertkov, G.~Falkovich, V.~Lebedev,
M.~Stepanov and M.~Vergassola
for numerous discussions.
I.~K. thanks U.~Frisch for the hospitality in Nice,
at the Observatoire de la C\^ote d'Azur,
where part of the work was done.
This work was supported in part by the Russian
Foundation for Basic Research (I.~K., grant 98-02-178/4) and
by the Italian M.U.R.S.T. (A.~G.).

\appendix

\section{Computation of the Jacobian}
In this appendix we derive the form of the Jacobian
term (\ref{jak}). Let us introduce a basis \( \hat {\rm e} _{ij} \) of the matrix algebra \( {\rm gl}(d) \) defined
by
\[
(\hat {\rm e} _{ij})_{kl}=\delta _{ik}\delta _{jl}\, ,\]
make use of the notation
\[
\hat {\rm t} _{ij}=\hat {\rm e} _{ij},\; j>i,\quad \hat {\rm d} _{k}=\hat {\rm e} _{kk},\; k=1,\ldots \, d,\quad \hat {\rm r} _{ij}=\hat {\rm e} _{ij}-\hat {\rm e} _{ji},\; j>i\, ,\]
and parametrize
\[
\hat{{\rm R}} ={\cal T} \exp \left( \int _{0}^{t}\phi _{ij}\hat {\rm r} _{ij}{\rm d} \tau \right) ,\quad \hat{{\rm D}} =\exp \left( \rho _{k}\hat {\rm d} _{k}\right) ,\quad \hat{{\rm T}} ^{-1}=\exp \left( X_{ij}\hat {\rm t} _{ij}\right) \, ,\]
where summation on repeated
indices is assumed and \( \rho _{d}=-(\rho _{1}+\cdots +\rho _{d-1}) \).

Under a variation \( (\delta \phi _{ij},\delta \rho _{k},\delta X_{ij}) \)
the
matrix \( \hat\sigma =\dot {\hat{{\cal W}} }\hat{{\cal W}} ^{-1} \) varies
as
\[
\delta \hat\sigma =\dot {\hat {\rm A} }+[\hat {\rm A} ,
\hat\sigma ]=\sigma _{ij}^{({\rm r})}\hat {\rm r} _{ij}+\sigma ^{({\rm d})}_{k}
\hat {\rm d} _{k}+\sigma _{ij}^{({\rm t})}\hat {\rm t} _{ij},\quad \hbox{with }{\rm A}=
\delta \hat{{\cal W}} \cdot \hat{{\cal W}}^{-1}
\]
The measure \( {\cal D}\hat\sigma  \) is
invariant under the global symmetries
 \( \hat{{\cal W}} \rightarrow \hat{{\cal R}} _{0}\hat{{\cal W}}  \) and \( \hat{{\cal W}} \rightarrow \hat{{\cal W}} \hat{\cal T}^{-1}_{0} \), so that the Jacobian

\[
{\cal J}=\left|\det
\left( \frac{\delta \hat\sigma }{\delta (\phi ,\rho ,X)}\right)\right| \]
does not depend on the \( \hat{{\rm R}}  \) and \( \hat{{\rm T}}  \) variables. Setting then \( \hat{{\rm R}} = {\bf 1} \), \( \hat{{\rm T}} ={\bf 1} \) \em after
\em having computed the variation \( \delta \hat \sigma  \) we find
\begin{eqnarray*}
\sigma _{k}^{({\rm d})} & = & \delta \dot\rho _{k}\, ,\\
\sigma _{ij}^{({\rm r})} & = & \delta \phi _{ij}+(\partial ^{-1}\delta \phi _{ij})(\dot\rho _{i}-\dot\rho _{j})\, ,\\
\sigma _{ij}^{({\rm t})} & = & -e^{\rho _{i}-\rho _{j}}\delta \dot X _{ij}-2(\partial ^{-1}\delta \phi _{ij})(\dot\rho _{i}-\dot\rho _{j})\, .
\end{eqnarray*}
The Jacobian matrix has
thus triangular form and can be computed with the help of the regularization
\begin{equation}
\label{regol}
\theta (0)\rightarrow \frac{1}{2},\quad \delta (0)=\frac{1}{h},\quad \delta '(0)=\frac{1}{h^{2}},\quad h\rightarrow 0\, .
\end{equation}
The result (\ref{jak}) is finally obtained making use of the simple identity
\[
\sum _{1\le i<j\le d}(\rho _{i}-\rho _{j})=\sum _{j=1}^{d}(d-2j+1)\rho _{j}=2\sum _{j=1}^{d-1}(d-j)\rho _{j}\, .\]
Note
that the consistency of the regularization prescription (\ref{regol}) can be
checked by computing \( \langle [\hat{{\cal W}} (t){\bf n} _{0}]^{2}\rangle  \) first directly and that using (\ref{jak}).



\begin{thebibliography}{99}

\bibitem{batchelor59}
G.~K. Batchelor, J. Fluid Mech. {\bf 5}, 113 (1959).

\bibitem{kraichnan74}
R.~H. Kraichnan, J. Fluid Mech. {\bf 64},  737  (1974).

\bibitem{ckv97P}
M. Chertkov, I. Kolokolov, and M. Vergassola, Phys.Rev.E, {\bf 56},
5483 (1997).

\bibitem{cfk98}
M. Chertkov, G. Falkovich, and I. Kolokolov,
Phys. Rev. Lett. {\bf 80}, 2121 (1998).

\bibitem{kolokolov86}
I.~V. Kolokolov, Phys. Lett. A {\bf 114},  99  (1986);
Ann. of Phys. {\bf 202},  165  (1990).

\bibitem{cfkl95}
M. Chertkov, G. Falkovich, I. Kolokolov, and V. Lebedev, Phys. Rev. E {\bf 51},
   5068  (1995).

\bibitem{fl94}
G. Falkovich and V. Lebedev, Phys. Rev. E {\bf 50}, 3883 (1994).

\bibitem{cgk94}
M. Chertkov, A. Gamba, and I. Kolokolov, Phys. Lett. A {\bf 192},  435  (1994).

\bibitem{gk96}
A. Gamba and I.~V. Kolokolov, J. Stat. Phys. {\bf 85},  489  (1996).

\bibitem{bgk97P}
D. Bernard, K. Gaw\c{e}dzki, and A. Kupiainen,
J. Stat. Phys. {\bf 90}, 519 (1998).

\bibitem{ChKVl}
M. Chertkov, I. Kolokolov  and M. Vergassola,
Phys. Rev. Lett. {\bf 80}, 512 (1998).

\bibitem{kls98P}
I.~Kolokolov, V.~Lebedev, and M.~Stepanov, in preparation.

\bibitem{kraichnan68}
R.~H. Kraichnan, Phys.~Fluids {\bf 11},  945  (1968).

\bibitem{ss94}
B.~I. Shraiman and E.~D. Siggia, Phys. Rev. E {\bf 49},  2912  (1994).

\bibitem{gr94}
I.~S. Gradshteyn and
I.~M. Ryzhik, {\em Table of integrals, series and products}
(Academic Press, San Diego, 1994).

\bibitem{BalF}
E.Balkovsky and G. Falkovich, Phys. Rev. E {\bf 57},
   1231  (1998).

\end{thebibliography}
\end{document}